\documentclass[prd,showpacs,preprintnumbers]{revtex4} 

\usepackage{graphicx}
\usepackage{dcolumn}
\usepackage{bm}
\usepackage{latexsym}
\usepackage{amsfonts}
\usepackage{amssymb}
\usepackage{amsmath}

\begin{document}

\preprint{KUNS-2125}

\title{
Circular Polarization of Primordial Gravitational Waves \\
in String-inspired Inflationary Cosmology
}

\author{Masaki Satoh}
\email{satoh@tap.scphys.kyoto-u.ac.jp}
\author{Sugumi Kanno}
\email{sugumi@hep.physics.mcgill.ca}
\author{Jiro Soda}
\email{jiro@tap.scphys.kyoto-u.ac.jp}
\affiliation{
 $^{\ast, \ddagger}$Department of Physics,  Kyoto University, Kyoto 606-8501, Japan\\
 $^\dagger$Department of Physics,  McGill University, Montr\'{e}al, QC H3A 2T8, Canada
}%

\date{\today}

\begin{abstract}
We study a mechanism to produce the circular polarization of
primordial gravitational waves. 
 The circular polarization is generated during the super-inflation
 driven by the Gauss-Bonnet term in the string-inspired cosmology. 
 The instability in the tensor mode caused by 
 the Gauss-Bonnet term and the parity violation due to the gravitational 
 Chern-Simons term are the essential ingredients of the mechanism. 
  We also discuss detectability of the produced circular polarization 
  of gravitational waves. It turns out that the simple model of single-field
 inflation contradicts CMB observations. To circumvent this difficulty, 
we propose a two-field inflation model.
In this two-field model,
  the circular polarization of gravitational waves is created in the 
  frequency range designed by the Big-Bang Observer (BBO) or
  the deci-hertz gravitational-wave observatory (DECIGO). 
\end{abstract}

\maketitle

\section{Introduction}

It is widely believed that the most promising candidate for an unified theory
including  quantum gravity is superstring theory.
It is therefore interesting to prove or disprove superstring theory
 from an observational point of view.   To this end, primordial
gravitational waves have been considered as the most efficient probe,
since the gravitational waves can carry the information 
of the universe at the Planck time.

In this paper, we focus on a robust prediction of superstring theory,
namely, the parity violation due to the gravitational Chern-Simons term.
In fact, the Chern-Simons term appears in the Green-Schwarz mechanism
which is necessary to cancel the anomaly in the theory~\cite{Green:1984sg}. 
It also arises as a string correction~\cite{Antoniadis:1992sa}.
Intriguingly, the existence of the Chern-Simons term 
can be probed by the primordial gravitational waves.
 This is because the Chern-Simons term can generate the circular polarization 
 of gravitational waves through the parity violation.
 While it is difficult to find other effects to produce the circular polarization
 of primordial gravitational waves. 
 Hence, if we detect the circular polarization
in the primordial gravitational waves, it would be a strong
indication of existence of the Chern-Simons term in the
early universe. Thus, it can be interpreted as an evidence of superstring theory.
In addition to this fundamental motivation, 
there is a phenomenological interest in Chern-Simons gravity
as an alternative theory of gravity~\cite{Jackiw:2003pm}.
Even from this phenomenological point of view, the circular polarization
of gravitational waves deserve detailed investigation.

Concerning the issue if the Chern-Simons term 
can produce the primordial gravitational waves with 
circular polarization, there has been already several works~\cite{Lue:1998mq,Choi:1999zy,Alexander:2006wk}.
It turned out that, however, there
exists no observable amount of circular polarization of gravitational
 waves~\cite{Alexander:2004wk,Lyth:2005jf}, 
if the background spacetime undergoes the standard slow-roll inflation.
The purpose of this paper is to point out that this disappointing result
 is reversed if we properly take into account another stringy effect. 
The point is that the  Chern-Simons term is not the only term that could be induced by the
stringy corrections. There is another term, the so-called 
Gauss-Bonnet term. If we consider the Chern-Simons term,  it would be
mandatory to incorporate the Gauss-Bonnet term. 
 Hence, we consider both terms, the Gauss-Bonnet and Chern-Simons terms, 
 and study a possibility to 
 have the circular polarization of primordial gravitational waves. 
 The reason we expect a different result from the one obtained from the standard 
  slow-roll inflation is the existence of the super-inflationary phase
 induced by the  Gauss-Bonnet 
 term~\cite{Antoniadis:1993jc,Kawai:1998bn,Kawai:1999xn,Kawai:1999pw,Cartier:2001is}. 
 And more importantly, as was firstly discovered
  in~\cite{Kawai:1997mf,Kawai:1998ab,Soda:1998tr}, 
 there exists an instability in gravitational wave modes
 during the super-inflationary stage. It is this instability that 
 generates the circular polarization of primordial gravitational waves.
 In fact, we show the primordial gravitational waves are fully polarized
due to the Gauss-Bonnet term. 
 We also discuss the detectability of the polarization of gravitational waves.
 The detectability of circular polarization of gravitational waves depends on when
the super-inflation occurs.
 Since the curvature perturbation has a blue spectrum in the super-inflationary
 regime, we need to assume the super-inflation occurs in the late stage of the
 slow-roll inflation. 
 As a realization, we propose a two-field inflation model
 where the circular polarization of gravitational waves could be
  produced in the BBO~\cite{bbo} or DECIGO~\cite{Seto:2001qf} frequency range.

The organization of this paper is as follows. 
In section II, we present basic equations. 
First, we study the evolution of the background spacetime
 in the Gauss-Bonnet-Chern-Simons gravity. 
For the background evolution, the Gauss-Bonnet term is crucial, 
while the Chern-Simons term plays no role. 
We show there are two types of inflation, namely, the standard slow-roll inflation 
and the super-inflation. Next, we derive equations for gravitational waves.
Here, the Chern-Simons term plays an important role. 
 In section III, we explain the essential mechanism to
 produce the circular polarization of gravitational waves. 
 We quantize the gravitational waves in the super-inflationary
  background and calculate the degree
of the circular polarization of gravitational waves. 
There, we will show that the significant circular polarization is created
due to the instability caused by the Gauss-Bonnet term. We note that
the model presented in this section is just for illustration.
More realistic modes are considered in the next section.
In section IV, we discuss the detectability of the circular polarization
generated during the super-inflationary regime.
We point out the defect of the single-field inflation model. 
Then,  as a  model consistent with current observations,
we propose a two-field inflation model where Gauss-Bonnet and Chern-Simons
term couple with the inflaton only in the second stage of the inflation. 
The final section is devoted to the conclusion.

\section{Basic equations in Gauss-Bonnet-Chern-Simons Gravity}

Before discussing primordial gravitational waves from the quantum fluctuation 
during the inflation, we need to reconsider the chaotic inflationary scenario 
driven by an inflaton $\phi$ with a potential $V(\phi)$. 
Although the Chern-Simons term does not alter the background geometry,
the Gauss-Bonnet term change the inflationary scenario at the early stage of the
 inflation. The change would be important for the primordial gravitational waves. 

We start with  the action motivated from string theory
 given by~\cite{Antoniadis:1993jc}
\begin{eqnarray}
 S = \frac{1}{2}\int{\rm d}^4x\sqrt{-g}R
 -\int{\rm d}^4x\sqrt{-g}\left[\frac{1}{2}\nabla^\mu\phi\nabla_\mu\phi
 +V(\phi)\right] 
 -\frac{1}{16}\int{\rm d}^4x\sqrt{-g} \xi (\phi)R^2_{\rm GB}
 +\frac{1}{16}
 \int{\rm d}^4x\sqrt{-g} \omega (\phi)R\tilde{R} \ ,
 \label{act0}
\end{eqnarray}
where the first term of action is the Einstein-Hilbert term and
$g$ is the determinant of the metric $g_{\mu\nu}$. 
We have set the unit $8\pi G=1$.
In the above action (\ref{act0}), we have taken into account the Gauss-Bonnet term 
\begin{eqnarray}
 R^2_{\rm GB} =R^{\alpha\beta\gamma\delta}R_{\alpha\beta\gamma\delta}
 -4R^{\alpha\beta}R_{\alpha\beta}+R^2 
\end{eqnarray}
and the Chern-Simons term 
\begin{eqnarray}
 R\tilde{R}&=\frac{1}{2}\epsilon^{\alpha\beta\gamma\delta}R_{\alpha\beta\rho\sigma}
 {R_{\gamma\delta}}^{\rho\sigma}  \ ,
\end{eqnarray}
where  $\epsilon^{\alpha\beta\gamma\delta}$ is the Levi-Civita tensor density.
We have also allowed the coupling of the inflaton field both to the 
Gauss-Bonnet $\xi (\phi)$ and Chern-Simons terms $\omega (\phi)$.
 Otherwise, these topological terms vanish identically. It should be noted that, 
as is well known, the Chern-Simons term does not contribute to the dynamics of 
the isotropic and homogeneous universe.

From now on, for simplicity,
we consider a typical potential, $V=1/2m^2\phi^2$.
We also specify the coupling function $\xi(\phi) =\omega(\phi)= 16 \phi^4$
for concreteness, although we keep them in the formula as far as we can. 
For other coupling functions, the analysis is similar and easy to perform.

\subsection{Background spacetime}

Let us consider the background spacetime with  
  spatial isotropy and  homogeneity. Then, the metric is given by 
\begin{align}
  ds^2 = g_{\mu\nu}{\rm d}x^\mu{\rm d}x^\nu =
 a^2 (\eta) \left[ - d\eta^2 + \delta_{ij}{\rm d}x^i{\rm d}x^j \right] \ .
\end{align}
Here, we have also assumed the flat space and used the conformal time $\eta$. 
Taking the variations of the action (\ref{act0}), we have equations
\begin{eqnarray}
 && 3{\cal H}^2 = \frac{1}{2}{\phi}^{\prime 2} +\frac{1}{2}m^2 a^2 \phi^2
 +\frac{3}{2a^2} {\cal H}^3 \xi' \label{eq1} \\
 && (2{\cal H}'+ 3 {\cal H}^2 )\left(1-\frac{1}{2a^2 }{\cal H} \xi' \right)
 +{\cal H}^2 \left(1+\frac{\cal H}{2a^2 }\xi' -\frac{1}{2a^2} \xi''\right)
 -m^2 a^2 \phi^2=0 \label{eq2} \\ 
 && \phi'' + 2 {\cal H} \phi' + \frac{3}{2a^2} {\cal H}^2{\cal H}' \xi_{,\phi}
 +m^2 a^2 \phi=0 
 \label{eom},
\end{eqnarray}
where we have defined ${\cal H}= a'/a$. Here, the prime denotes the derivative
with respect to the conformal time $\eta$. 

\begin{figure}[htbp]
 \centering
 \includegraphics[scale=0.8]{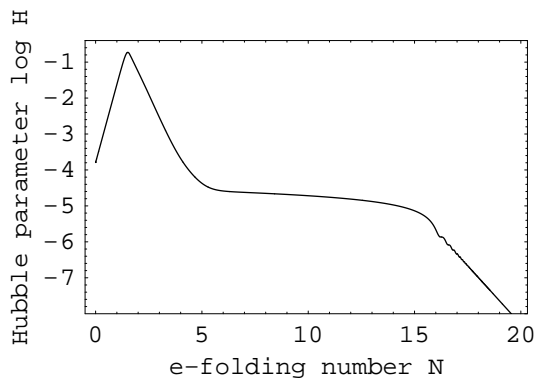}
 \caption{A typical evolution of the background spacetime is numerically calculated
 and  displayed.
 A short period of the super-inflationary phase is followed by 
 a long period of the slow-roll inflationary phase. }
 \label{evol}
\end{figure}

\begin{figure}[htbp]
 \centering
 \includegraphics[scale=0.8]{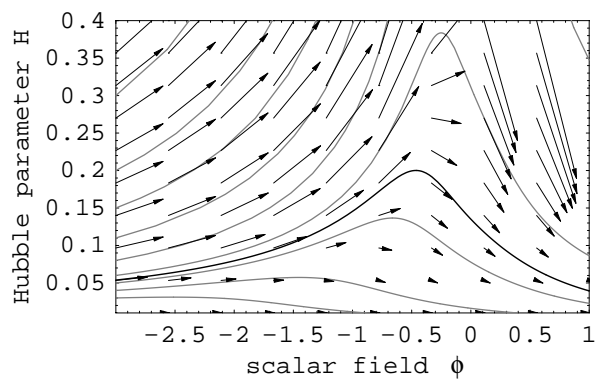}
 \caption{Phase diagram around $\phi\simeq 0$ is drawn. The thick line in the diagram 
 denotes the trajectory
 corresponding to Fig. \ref{evol}. This tells that the asymptotic solution breaks down
 before reaching $\phi=0$, and thus the singularity is avoided.}
 \label{phase}
\end{figure}

In the conventional case, namely, without the Gauss-Bonnet term,
the Hubble friction in the equation of motion for the scalar field
makes the rolling of the scalar field slow. Then, the slow-roll
inflation is soon realized for the initial condition with a large Hubble.
 However, in the presence of the Gauss-Bonnet term, the force
due to the Gauss-Bonnet term becomes dominant for a large value of $\phi$.
If the scalar field start with a negative value, the force
term accelerate the scalar field and makes the kinetic term dominant.
Hence,  the situation, 
${\cal H}^2\ll  {\phi}^{\prime 2}, m^2 a^2 \phi^2\ll {\phi}^{\prime 2}$, is realized.
 Thus, Eqs.(\ref{eq1}) and (\ref{eq2}) become
\begin{eqnarray}
 && a^2 {\phi}^{\prime 2}  +  3 {\cal H}^3 \xi' = 0 \label{super1} \\
 &&  (2{\cal H}'+ 3 {\cal H}^2 ) \xi' 
 +{\cal H} \left(   \xi'' - {\cal H}\xi' \right) =0 \ .
 \label{super2}
\end{eqnarray}
The scalar field $\phi$ rolls down from the negative side towards zero
according to the above equations (\ref{super1})  and (\ref{super2}).
Now, it is easy to solve Eqs.(\ref{super1}) and (\ref{super2}) as
\begin{eqnarray}
 \phi=- \sqrt{15/16} (-\eta)^{5/6} \ , \quad
 a(\eta ) = (-\eta)^{-1/6} \ , \quad
 {\cal H} = \frac{1}{(- 6\eta)} \ , \quad \eta <0 \ .
  \label{assum}
\end{eqnarray}
We note that the super-inflation $H'  >0$ is realized in this phase.
Here, $H= {\cal H}/a$ is the Hubble parameter.  
As the scalar field rolls down, the Gauss-Bonnet term decreases.
Eventually, the conventional Hubble friction overcome 
the Gauss-Bonnet effect and the slow-roll inflation commences.
A typical evolution of the spacetime is shown in Fig.\ref{evol}.
In the super-inflationary phase $H' >0$,  the weak energy condition
is violated. Hence, the system may show the instability.
Of course, as you can see in Fig.~1, this instability is a transient one.
To make this point clear, we rewrite the equations of motion (\ref{eom}) as the
autonomous system for $\phi$ and $H$ and plot the phase flow diagram    
in Fig.~\ref{phase}.   
This diagram shows that the super-inflationary phase will be followed by
the standard phase where the Hubble parameter is decreasing. 

In the next section, we will show that
the transient super-inflation phase gives rise to the circular polarization of gravitational waves.

\subsection{Action for  gravitational waves}

In this subsection, we will deduce the quadratic action for the gravitational
waves from the action (\ref{act0}). 
Let us consider the tensor perturbation
\begin{align}
 {\rm d}s^2=g_{\mu\mu}{\rm d}x^\mu {\rm d}x^\nu
 =a^2(\eta)\left[-{\rm d}\eta^2
 +(\delta_{ij}+h_{ij}){\rm d}x^i{\rm d}x^j\right]  \ ,
\end{align}
where  $h_{ij}$ satisfies the transverse-traceless conditions $h_{ij,j}=h_{ii}=0$.
Substituting the metric into the Einstein action with the Gauss-Bonnet term,
we obtain the quadratic action~\cite{Soda:1998tr,Gasperini:1997up,Cartier:2001gc}
\begin{eqnarray}
&& \frac{1}{2}\int{\rm d}^4x \sqrt{-g}R
 -\int{\rm d}^4x\sqrt{-g}\left[\frac{1}{2}\nabla^\mu\phi\nabla_\mu\phi
 +V(\phi)\right] -\frac{1}{16}\int{\rm d}^4x \sqrt{-g}
             \xi (\phi) R^2_{\rm GB}  \nonumber\\
&& \qquad = \frac{1}{8}
 \int{\rm d}^4x a^2\left[ \left(1- \frac{{\cal H}\xi'}{2a^2} \right) h'^{ij}h'_{ij}
 - \left( 1+ \frac{{\cal H}\xi'}{2a^2} - \frac{\xi''}{2a^2} \right) h^{ij|k}h_{ij|k}
       \right]  \ .
\end{eqnarray}
In addition to these terms, the parity violating Chern-Simons term gives
\begin{eqnarray}
\frac{1}{16}
 \int{\rm d}^4x \sqrt{-g} \omega (\phi) R\tilde{R}
 =-\frac{1}{16}\int{\rm d}^4x \omega' \epsilon^{ijk}
 \left(h'_{ai}{h'^a}_{k|j}-h_{ai|b}{{{h^a}_k}^{|b}}_{|j}\right),
\end{eqnarray}
where the stem $|$ denotes a derivative with respect to the spatial coordinates. 
Here, we have used the convention $\epsilon^{0ijk}=-\epsilon^{ijk}$ with
$\epsilon^{123}\equiv 1$. It is convenient to expand $h_{ij}$ by plane waves
\begin{align}
 \frac{h_{ij}(\eta,{\bf x})}{\sqrt{2}}=
 \sum_{A={\rm +,\times}}\int\frac{{\rm d}^3k}{(2\pi)^3}\psi^A_{\bf k}(\eta)
 {\rm e}^{i{\bf k}\cdot{\bf x}}p^A_{ij},
\end{align}
where we have introduced the circular polarization tensor $p^A_{ij}$ ($A=R,L$)
defined by
\begin{align}
p^{\rm R}_{ij}\equiv\frac{1}{\sqrt{2}}
\left(p^{ +}_{ij}+ip^{\times}_{ij}\right),
\quad
p^{\rm L}_{ij}\equiv\frac{1}{\sqrt{2}}
\left(p^{ +}_{ij}-ip^{\times}_{ij}\right) \ .
\end{align}
Here, the polarization tensors $ p^{ +}_{ij}$ and $p^{\times}_{ij}$
are the plus and cross modes, respectively. 
The polarization tensors are normalized as 
$p^{*A}_{ij} p^B_{ij} =2 \delta^{AB}$. 
The circular polarization has the property
\begin{align}
\frac{k_s}{k}\epsilon^{rsj}p^A_{ij}=-i\lambda^A({p^r}_i)^A,
\quad A={\rm R},{\rm L},
\end{align}
where $\lambda^{\rm R}\equiv+1$ and $\lambda^{\rm L}\equiv-1$
 represent the right and left-handed circular polarizations, respectively. 
This makes the action of the Chern-Simons term diagonal.

Putting  above results together  and defining 
a new variable $\mu_{\bf k}^A \equiv z_{\bf k}^A \psi^A_{\bf k}$, we obtain
\begin{align}
 S^{\rm GW} = \sum_{A={\rm R,L}}\frac{1}{2}
 \int{\rm d}\eta\frac{{\rm d}^3k}{(2\pi)^3}
 \left[|\mu_{\bf k}^{\prime A}|^2-\left(1+\frac{{\cal H} \xi' }{z_A^2} 
 -\frac{\xi''}{2z_A^2}\right) k^2 |\mu_{\bf k}^A|^2
 +\frac{z''_A}{z_A} |\mu_{\bf k}^A|^2\right],
\end{align}
where we have defined 
\begin{align}
 z_A = a(\eta)\sqrt{1-\frac{{\cal H} \xi'}{2 a^2}
                 -\lambda^A k\frac{\omega' }{2 a^2}}  \ ,
 \qquad \lambda^{\rm R}=1 \ ,    \quad \lambda^{\rm L}=-1, \label{za}  \ .
\end{align}
Thus, the equation of motion becomes
\begin{align}
 (\mu_{\bf k}^A)''+\left[\left(1+ \frac{{\cal H}\xi'}{z_A^2} -\frac{\xi''}{2 z_A^2}\right)
 k^2-\frac{z''_A}{z_A}\right]\mu_{\bf k}^A=0.
 \label{eom_gw}
\end{align}
The term $z''_A/z_A$ can be interpreted as the effective potential. 
The coefficient of the wavenumber $k$ may be interpreted as the square 
of the speed of sound waves.

In the long wavelength limit, we have the solution
\begin{eqnarray}
  \mu_{\bf k}^A  
  = G z_{\bf k}^A + D z_{\bf k}^A \int \frac{d\eta}{z_{\bf k}^{A2}}  \ ,
\end{eqnarray}
where $G,D$ are the constants of integration.
In the conventional inflationary scenario, 
the first term corresponds to the growing mode which is actually 
constant when we translated back to the metric perturbations.
Hence, the primordial gravitational waves are frozen on super-horizon
scales. However, in the presence of Gauss-Bonnet and Chern-Simons
terms, the role of the growing and decaying modes are interchanged.
This gives rise to an interesting effect on the polarization. 

\section{ A mechanism to produce Circular Polarization}

In this section, we present a mechanism to produce the circular polarization
of primordial gravitational waves. The interplay between the instability
induced by the Gauss-Bonnet term and the parity violation due to Chern-Simons
term is essential for the mechanism. The mechanism is efficient and
generic in the sense that both terms are ubiquitous in string theory.

Now, let us quantize the gravitational waves and calculate the degree of
polarization. 
What we should do is to promote $\mu_{\bf k}^A$ to the operator and
expand by mode functions $u_{\bf k}^A (\eta)$ as 
\begin{eqnarray}
  \mu_{\bf k}^A  
  = a_{\bf k}^A u_{\bf k}^A + a_{\bf k}^{\dagger A} u_{\bf k}^{*A} \ ,
\end{eqnarray}
where $u_{\bf k}^{*A}$ is the complex conjugate of $u_{\bf k}^{A}$
and  $a_{\bf k}^A$ and $a_{\bf k}^{\dagger A}$
 denote the annihilation and creation operators.
The mode functions $u_{\bf k}^{A}$ satisfies the same equation 
for  $\mu_{\bf k}^A$.   
Once the positive frequency mode $u_{\bf k}^{A}$ is specified, 
the vacuum is define by
\begin{eqnarray}
 a_{\bf k}^A |0> =0 \ .
\end{eqnarray}
 Then, we can calculate the vacuum fluctuations
\begin{eqnarray}
 <0|  |\mu_{\bf k}^A|^2 |0> =  |u_{\bf k}^A|^2 \ .
\end{eqnarray}
Here, we focus only on the gravitational waves which leaves the
horizon during super-inflationary phase, because the gravitational waves
still under the horizon when super-inflation ends are
 the same as those of the slow-roll inflation.  

During the stage prior to the slow-roll inflation, 
 $z_A$ can be approximated as
$
 z_A=\sqrt{-{\cal H}\xi'/2 - \lambda^A k \omega' /2 }.
$
In spite of this simplification, it is still  difficult to 
solve the equation of motion analytically. 
Hence, we solve the equation at super-horizon scale and sub-horizon
scale, separately. After that, we
  connect these solutions smoothly at the horizon crossing.

First, let us consider the sub-horizon scale,  
${\cal H}\ll k$ or $ -k\eta\gg 1/6$.  
Leaving the terms up to the order of $1/(-k\eta)$, 
we obtain the equation  
\begin{align}
 (u_{\bf k}^A)''+k^2\left(1-\lambda^A\frac{8}{3}\frac{1}{-k\eta}
 \right) u_{\bf k}^A=0  \ .
\end{align}
This can be solved analytically by using confluent hypergeometric functions.
Instead of doing that, we estimate the degree of the polarization approximately. 
The initial condition can be set by imposing the Bunch-Davies vacuum.
More precisely, we choose the positive frequency mode
\begin{align}
 u^A_{\bf k}=\frac{1}{\sqrt{2k}}{\rm e}^{-ik\eta}
\end{align}
in the asymptotic past. Namely, deep inside  the Hubble horizon,
 both the left and right polarization modes are simply
 oscillating.  At the time $- k\eta = 8/3$, the asymmetry shows up.
After that time,  $-k\eta < 8/3$, we can approximate the solutions as
\begin{align}
 u^A_{\bf k}= A_1 \left(\frac{3\lambda^A (-k\eta)}{8 k^2 } \right)^{1/4}
                  \exp \left( 2\sqrt{\frac{8\lambda^A }{3}(-k\eta) } \right) 
                  + A_2 \left(\frac{3\lambda^A (-k\eta)}{8 k^2 } \right)^{1/4}
                  \exp \left( - 2 \sqrt{\frac{8\lambda^A }{3}(-k\eta) } \right)\ ,
\end{align}
where we have used WKB approximation which is valid for the
period we are considering. 
To determine  the constants of integration $A_1$ and $A_2$, 
we have to match solutions smoothly at $- k\eta \sim 8/3$ as
\begin{align}
 \left.u_{\bf k}^A(\eta)\right|_{\eta < 8/3}=
 \left.u_{\bf k}^A(\eta)\right|_{\eta > 8/3 },
 \quad
 (\left.u_{\bf k}^A)'(\eta)\right|_{\eta < 8/3}=
 (\left.u_{\bf k}^A)'(\eta)\right|_{\eta > 8/3}.
\end{align}
Apparently, both $A_1 $ and $A_2$ are not zero. Hence, 
for the left-handed circular polarization modes $\lambda^L =-1$, 
the solution shows damping oscillation.
While, for the right-handed circular polarization mode $\lambda^R =1$, 
the solution  grows rapidly. 
It is this instability  due to the Gauss-Bonnet term that produces the
 difference between  the left and right-handed circular polarization modes. 
The ratio can be quantified by
\begin{eqnarray}
  \frac{|u_{\rm R} |^2}{|u_{\rm L}|^2} = \frac{\exp(- 8/3 )}{\exp(- 32/3)}
  \sim 2980   \ . 
\end{eqnarray}
Thus, we can expect the fully polarized gravitational waves. 

To make the analysis more accurate, 
we need the solutions to  continue to the super-horizon scale,
${\cal H}\gg k$ or $ -k\eta\ll 1/6$. 
On the super-horizon scales, the solution can be obtained as
\begin{eqnarray}
 (u_{\bf k}^A)''+\frac{2}{9}k^2\frac{1}{(-k\eta)^2}u_{\bf k}^A=0.
\end{eqnarray}
This can be solved easily as
\begin{eqnarray}
 u_{\bf k}^A=B_1(-k\eta)^{1/3}+B_2(-k\eta)^{2/3}  \ ,
\end{eqnarray}
where $B_1$ and $B_2$ are  constants of integration. 
To determine  the constants $B_1$ and $B_2$, we have to match solutions smoothly
at $-\eta_{\rm cross}=1/(6k)$ as
\begin{align}
 \left.u_{\bf k}^A(\eta_{\rm cross})\right|_{{\cal H}\ll k}=
 \left.u_{\bf k}^A(\eta_{\rm cross})\right|_{{\cal H}\gg k},
 \quad
 (\left.u_{\bf k}^A)'(\eta_{\rm cross})\right|_{{\cal H}\ll k}=
 (\left.u_{\bf k}^A)'(\eta_{\rm cross})\right|_{{\cal H}\gg k}.
\end{align}
\begin{figure}[htbp]
 \centering
 \includegraphics[scale=0.8]{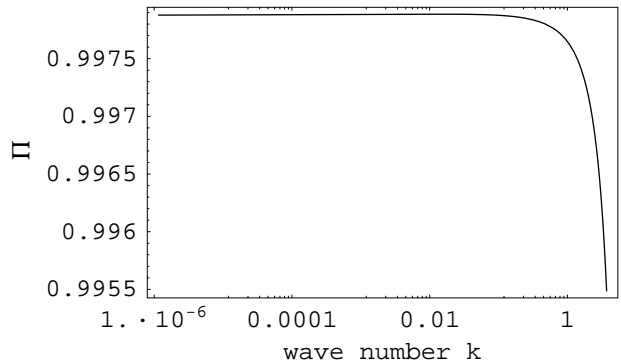}
 \caption{The degree of the polarization $\Pi (k)$
 as a function of wave numbers $k$ is shown.
 As expected, the gravitational waves are almost 100\% circularly polarized.}
 \label{eta}
\end{figure}
What we want to calculate is the degree of the circular polarization 
defined as the difference between the power of right and left-handed  
circularly polarized gravitational waves at the end of the
super inflation phase:
\begin{align}
 \Pi (k)\equiv
 \frac{|u_{\bf k}^{\rm R}(\eta_{\rm end})|^2-|u_{\bf k}^{\rm L}(\eta_{\rm end})|^2}
 {|u_{\bf k}^{\rm R}(\eta_{\rm end})|^2+|u_{\bf k}^{\rm
 L}(\eta_{\rm end})|^2} \ ,
\end{align}
where $\eta_{\rm end}$ represents the time when the super-inflation ends.
We have performed  matching analysis and displayed a 
numerical result in Fig.\ref{eta}.  
Because of the behavior just before the horizon crossing, namely right-handed 
circular polarization mode is growing and
 left-handed circular polarization  mode is decaying, 
 consequently the net polarization is produced. 
The resultant polarization is sufficiently large and hence detectable by BBO 
or DECIGO.

To calculate the spectrum of the polarization of gravitational waves
 which is directly observed today, 
we have to connect solutions from super inflation
phase to the slow-roll inflation, radiation
dominant, and matter dominant phase. 
 This part of calculations is standard and does not alter the polarization.

Finally, we should mention a subtle point in the calculation.
From the definition of $z_A$ (\ref{za}), 
it is obvious that there is a point $\eta_{\rm sing}$ where $z_A=0$.
That occurs close to the time 
when the left-handed circular polarization mode cross the horizon. 
There, the solution behaves as
\begin{eqnarray}
u^{\rm L}_{\bf k}
=C_1\sqrt{|\eta-\eta_{\rm sing}|}
+C_2\sqrt{|\eta-\eta_{\rm sing}|}\ln(|\eta-\eta_{\rm sing}|) \ ,
\end{eqnarray}
where $C_1$ and $C_2$ are  constants of integration. 
We can see that $u^{\rm L}_{\bf k}\rightarrow 0$, in the limit 
$\eta\rightarrow\eta_{\rm sing}$,
 so it causes no problem to solve the Eq.(\ref{eom_gw}). 
 However, the relevant quantity
  is the physical amplitude $\psi^{\rm L}_{\bf k}$, rather than 
 $u^{\rm L}_{\bf k}$, namely,
\begin{align}
\psi^{\rm L}_{\bf k}=\frac{u^{\rm L}_{\bf k}}{z_{\rm L}}
\sim C_1 + C_2 \ln(|\eta-\eta_{\rm sing}|) \ .
\end{align}
Apparently, the physical amplitude diverges at $\eta\rightarrow\eta_{\rm sing}$. 
In the previous work~\cite{Alexander:2004wk,Lyth:2005jf}, 
this issue was serious, because this divergence means the breakdown of the
linear analysis. It suggests that we should take into account the non-linear
effect which is not available. 
In our case, the relevant mode is the right polarization mode
which has no divergence problem. 
As the breakdown of the linear analysis occurs just at the horizon
crossing, the non-linear effects works almost instantaneously. 
Hence, it is reasonable to assume 
that the non-linear effect for the left-handed circular polarization mode
 hardly affects the evolution of the right-handed circular polarization mode. 
 Therefore, we believe the consequence about the degree of polarization 
 does not change.

\section{Detectability of Circular Polarization of Gravitational waves}

First of all, we should emphasize that the analysis done in the previous 
section is explanatory. The frequency range where the circular polarization
is created will change depending  on
when the super-inflation occurs. In this section, we will discuss
the indirect detection through CMB 
and the direct detections by interferometer.

As to the in-direct detection of circular polarization through the
cosmic microwave background radiation (CMB),
the required degree of circular polarization has been obtained
as $|\Pi| \gtrsim 0.35 (r/0.05)^{-0.6}$~\cite{Saito:2007kt}, 
where $r$ is the tensor-to-scalar ratio. 
The relevant frequency in this case is around $f \sim 10^{-17}$ Hz. 
As to the direct detection of circular polarization,
the required degree of circular polarization has been estimated 
as $\Pi \sim 0.08 (\Omega_{\rm GW} /10^{-15})^{-1} ({\rm SNR}/5)$
around the frequency $f \sim 1$ Hz~\cite{Seto:2006hf,Seto:2006dz}, where
$\Omega_{\rm GW}$ is the density parameter of the stochastic gravitational waves
and SNR is the signal to the noise ratio. Here, 10 years observational time is assumed.
Of course, for the isotropic component of circular polarization
of stochastic gravitational waves, one needs to break the symmetry of the
detector configuration.

\begin{figure}[htbp]
 \centering
 \includegraphics[scale=0.8]{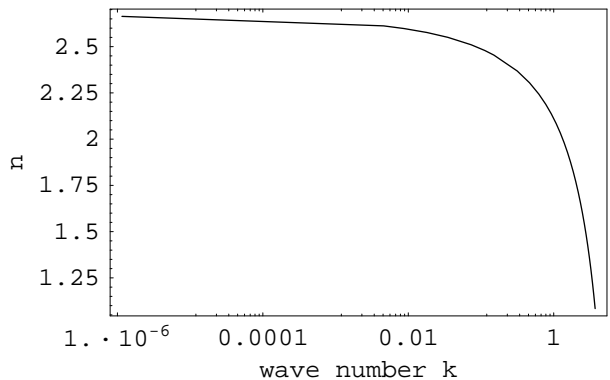}
 \caption{The numerical result of the 
 spectral index $n$ for the power spectrum of primordial 
 gravitational waves as a function of wavenumbers $k$ is shown. 
 In the relevant frequency range, we have obtained the spectral index $n = 2.66$
 numerically. }
 \label{spectrum}
\end{figure}
Thus, in order to know if we can detect polarized gravitational waves,
we need the degree of the polarization, 
the intensity of the gravitational waves $\Omega_{\rm GW}$ or 
the tensor-to-scalar ratio $r$,
and also the spectrum around the relevant frequency ranges. 
As we have shown already, we can get the polarization $\Pi \sim 1$ easily. 
Moreover, it is not difficult to get the required  $\Omega_{\rm GW}$
or $r$ provided standard model parameters.  
We can calculate the spectrum numerically assuming the Bunch-Davis vacuum. 
We have presented a numerical result in Fig.\ref{spectrum}. 
This result can be understood analytically.
In fact, the initial mode function behaves $\propto 1/\sqrt{k}$ inside the horizon
which should match to the growing mode $B_1 (-k\eta)^{1/3}$
 on the super horizon scales.
Matching condition at $-k\eta \sim1$ gives the $k$ dependence of
the coefficient $B_1$ as $B_1 \propto 1/\sqrt{k}$. 
Thus, we have $u_k \propto k^{1/3-1/2} \sim k^{-1/6}$. 
Consequently, the power spectrum for the gravitational waves
becomes $k^3 |u_k|^2 \propto k^{8/3}$. 
Since the spectrum has the blue index $n\sim 2.66$, the polarization
would be detectable even in the low intensity case. 
Now, the point is the frequency range, namely,
 the time when the polarization is created. 
In the following discussion, we will mainly focus on this issue.

In the case of the single inflaton model we have discussed, the frequency 
where the circular polarization is significant lies around the
CMB scale $ f \sim 10^{-17}$ Hz.  Hence, there is a possibility 
for the circular polarization to be indirectly detected 
through the CMB observation, more precisely, through the Temperature and $B$ mode 
or $E$ mode and $B$ mode correlations~\cite{Lue:1998mq,Saito:2007kt,Kahniashvili:2004}. 
In a recent paper~\cite{Saito:2007kt}, 
it is argued that a high degree of polarization is necessary
for the polarization to be detected in the future observations. 
The high degree of polarization can be attainable by the mechanism
we have found in this paper.
However, in this simple model, we have to worry about
the curvature perturbations which is also expected to have the blue spectrum
during the super-inflation. One possible way out of this problem
is to assume the curvature perturbation created by the inflaton is negligible and 
to resort to the curvaton scenario~\cite{Enqvist:2001zp,Lyth:2001nq,Moroi:2001ct}
 as is common to the pre-big-bang type
models~\cite{Lidsey:1999mc,Gasperini:2002bn,Kanno:2002py,Kanno:2005vq}.
However, it is not clear if the concrete realization of this kind of model
is possible. Moreover, one may feel that introducing the curvaton mechanism 
into the scenario is not appealing even if it exists.   
Fortunately, there exists another interesting possibility.
We will devote the remaining part of this section to the model
which is consistent with observational constraints.

To circumvent the problem of the curvature perturbations, 
 we propose a two-field inflation model defined by the action 
\begin{eqnarray}
S&=& \frac{1}{2}\int{\rm d}^4x\sqrt{-g}R
 -\int{\rm d}^4x\sqrt{-g}\left[\frac{1}{2}\nabla^\mu\chi\nabla_\mu\chi
 + \frac{1}{2}\nabla^\mu\phi\nabla_\mu\phi
 +V(\phi , \chi )\right] \nonumber\\
 && \qquad \qquad 
 -\frac{1}{16}\int{\rm d}^4x\sqrt{-g} \xi (\phi)R^2_{\rm GB}
 +\frac{1}{16}
 \int{\rm d}^4x\sqrt{-g} \omega (\phi)R\tilde{R} \ ,
\end{eqnarray}
where the potential $V(\phi ,\chi)$ is assumed to take a hybrid type.
The main difference between our model and the conventional inflation model is simply
the Gauss-Bonnet and Chern-Simons term. 
In the first stage, the field $\chi$ drives the
slow-roll inflation. During this stage, the field $\phi$ takes a constant large field value.
Hence, both the Gauss-Bonnet and the Chern-Simons terms are decoupled from the system.
During this conventional inflationary period, the temperature fluctuations
of CMB are created. And the spectrum of the fluctuations is almost flat. 
Then, at some point, the motion in the direction of $\phi$ is triggered and
the Gauss-Bonnet and Chern-Simons terms take part in the dynamics. 
In this second stage, the $\phi$ field rolls down the hill rapidly
and the Gauss-Bonnet term induces a super-inflationary phase. 
During this stage, the circular polarization of the primordial gravitational waves
 is produced through the parity violation of the  Chern-Simons term.
Subsequently, 
the second standard slow-roll inflation driven by the same field $\phi$ takes over. 

Here, we present an example.  
\begin{figure}[htbp]
 \centering
 \includegraphics[scale=0.8]{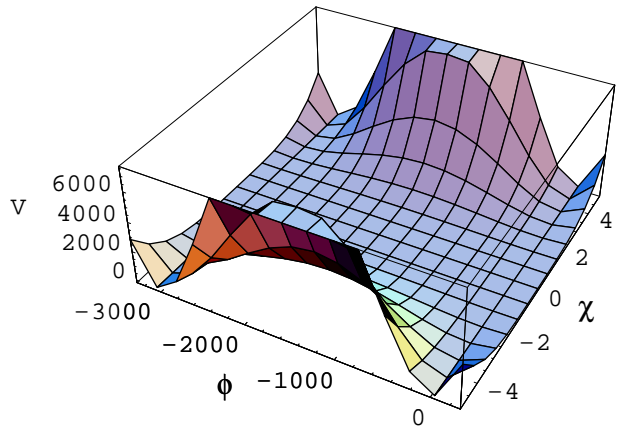}
 \caption{Potential function for the two field model is depicted.
  Initially, the scalar field $\phi$ is sticked to the point $\phi=-3000$ and
   the other scalar field $\chi$ slowly rolls down to give rise to the inflation. 
   At some point, $\phi$ starts to roll down toward $\phi=0$.
   Then, the coupling to the Gauss-Bonnet term induces the super-inflation 
   where the circular polarization of gravitational waves is created.
   Subsequently, the standard slow-roll inflation follows.}
 \label{v}
\end{figure}
The  potential function is
\begin{align}
 V(\phi,\chi)=  V_{\rm p}(\lambda_\chi \chi^4 + 1)
 \left\{
 \left(2/\phi_{\rm m}\right)^2 
 \left[(\phi + \phi_{\rm m}/2)^2 - (\phi_{\rm m}/2)^2\right]\right\}^2
 +\frac{1}{2}m_\chi^2  \chi^2 + \left(\frac{\phi}{\phi_{\rm m}}\right)^2 V_{\rm m}\ ,
\end{align}
where the parameters are set as 
\begin{align}
 \phi_{\rm m}=3000,\ V_{\rm m}=10^{-8},\ V_{\rm p}=0.235,
 \ \lambda_\chi=10^2,\ m_\chi=10^{-6}\  \ .
\end{align}
This potential is shown in Fig. \ref{v}.
Coupling $\xi$ and $\omega$ are same as those of the single-field model, namely 
$\xi=\omega=16\phi^4$.  We have calculated the evolution of Hubble 
parameter numerically (see Fig. \ref{evol2}). 
\begin{figure}[htbp]
 \centering
 \includegraphics[scale=0.8]{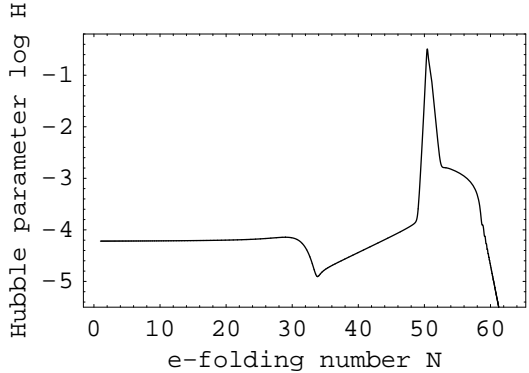}
 \caption{The time evolution of Hubble parameter is numerically calculated.
  The curvature perturbations on the CMB scales are produced during the slow-roll phase. 
  Hence, this model is consistent with CMB observations.
 Gauss-Bonnet term induces the super-inflation around $N\simeq 50$
 where the circular polarization would be created.   It could be observed
 directly through the  interferometer detector.}
 \label{evol2}
\end{figure}
\begin{figure}[htbp]
 \centering
 \includegraphics[scale=0.5]{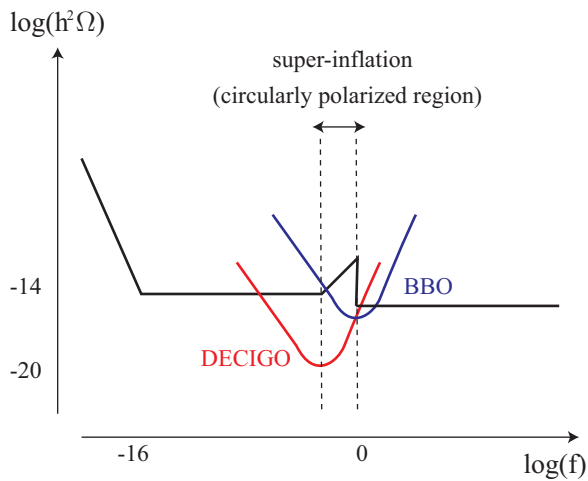}
 \caption{A schematic picture of the expected spectrum is depicted.
 The frequency range where we can observe the circular polarization
 depends on the model parameters. In that range, the spectrum is blue. 
 In other parts, the spectrum is almost flat. 
 The expected sensitivities of BBO and DECIGO are also plotted.}
 \label{picture}
\end{figure}

For this model, the duration of the first and second slow-roll inflation is tunable.
Hence, it is possible to have the gravitational waves with the significant circular
polarization in the BBO or DECIGO frequency range (see Fig.\ref{picture}). 
Since the spectrum is blue, even the ground based detectors 
may find the circular polarization as is pointed out recently~\cite{taruya}. 

To make a more precise prediction of the polarization of gravitational waves, 
we need more complete analysis. In particular, we should calculate the spectrum 
of the curvature perturbations of this model.  
We leave these details for future publications~\cite{satoh}. 

\section{Conclusion}

 We have studied a mechanism to produce the circular polarization of 
 gravitational waves in the string-inspired cosmology.
 It turned out that the circularly polarized gravitational waves
 are ubiquitous in string cosmology. 
 The point we have observed is that there are two key terms in string theory, 
 namely, the Chern-Simons term and the Gauss-Bonnet term.
  The Chern-Simons term violates the parity invariance, therefore
  it makes a room for the circularly polarized gravitational waves
  to be produced.  However, in the previous 
  works~\cite{Lue:1998mq,Alexander:2004wk,Lyth:2005jf}, it had been shown
   that there is no significant circular polarization of gravitational waves 
   within the conventional inflationary scenario.
   In this paper, we have shown that the Gauss-Bonnet term reversed the previous
   conclusion.   The Gauss-Bonnet term  has changed the background evolution 
   in such a way that the super-inflationary epoch appears 
   during the conventional inflationary stage. During the super-inflation, 
   there exists an instability in the tensor modes.
   It is the instability that produces a significant circular polarization 
   $\Pi \sim 1$. 
   We have also discussed the detectability of the polarization. 
   We have shown that the detectability depends on the specific scenario
   of inflation. In the single-field  inflation model, the circular polarization
   would be observable at the CMB scale. The consistency with the observation of
   the temperature fluctuations requires the curvaton scenario in this case.
   To circumvent the situation, 
   we have also proposed a two-field inflation model.
   In the two-field model,  
   the frequency range where the circular polarization is produced
   depends on the parameter. It would be possible to make a model to 
   produce gravitational waves with circular polarization
   in the  BBO or DECIGO frequency range.
   In this case,  curvature fluctuations on large scales are produced
   by the standard slow-roll inflation. Hence, we do not need the curvaton scenario
   any more. 

There are several directions to be explored. 
We should investigate the two-field inflation model in detail~\cite{satoh}. In particular,
it is interesting to seek a concrete realization of the model
in the string landscape. 
It is also intriguing to search other mechanism to produce a circular polarization
of gravitational waves. 
One simple possibility is to consider the non-vacuum initial state
in Lorentz violating inflationary scenario~\cite{Kanno:2006ty}.
The coupling to the Lorentz violating sector could violate the parity invariance. 
Hence, it is possible to take the asymmetric vacuum for right and left-handed circular 
polarizations. In fact, the Chern-Simons term already allows this possibility.
There may be other interesting mechanisms with non-inflationary origin,
such as the turbulence~\cite{Kahniashvili:2005qi}
or the helical magnetic fields~\cite{Kahniashvili:2004,Kahniashvili:2006}. 
The Gauss-Bonnet term at present may become an interesting subject
 in the light of the dark 
 energy~\cite{Calcagni:2005im,Koivisto:2006ai,neupane06a,neupane06b,Leith:2007bu,neupane07,Nojiri:2007te,Elizalde:2007pi}.

\vskip 2cm
J.S. is supported by  
the Japan-U.K. Research Cooperative Program, the Japan-France Research
Cooperative Program,  Grant-in-Aid for  Scientific
Research Fund of the Ministry of Education, Science and Culture of Japan 
 No.18540262 and No.17340075.  
S.K. is supported by JSPS Grant-in-Aid 
for Research Abroad.

\end{document}